\documentclass{article}
\usepackage{graphicx}
\usepackage[numbers]{natbib}
\usepackage[dvipsnames]{xcolor}
\usepackage{float}

\usepackage[preprint]{neurips_2023}

\usepackage[utf8]{inputenc} 
\usepackage[T1]{fontenc}    
\usepackage{url}            
\usepackage{booktabs}       
\usepackage{amsfonts}     
\usepackage{nicefrac} 
\usepackage{microtype} 
\usepackage{xcolor}

\title{Hopfield-Enhanced Deep Neural Networks for Artifact-Resilient Brain State Decoding}

\author{%
  Arnau Marin-Llobet\\
  Harvard University \& FRCB-IDIBAPS\\
  Allston, USA \\
  \texttt{amarinllobet@seas.harvard.edu} \\
  \And
  Arnau Manasanch\\\
  FRCB-IDIBAPS\\
  Barcelona, Spain \\
  \texttt{manasanch@recerca.clinic.cat} \\
  \And
  Maria V. Sanchez-Vives \\
  FRCB-IDIBAPS \& ICREA \\
  Barcelona, Spain \\
  \texttt{msanche3@recerca.clinic.cat} \\
}

\begin{document}

\maketitle

\begin{abstract}
The study of brain states, ranging from highly synchronous to asynchronous neuronal patterns like the sleep-wake cycle, is fundamental for assessing the brain's spatiotemporal dynamics and their close connection to behavior. However, the development of new techniques to accurately identify them still remains a challenge, as these are often compromised by the presence of noise, artifacts, and suboptimal recording quality. In this study, we propose a two-stage computational framework combining Hopfield Networks for artifact data preprocessing with Convolutional Neural Networks (CNNs) for classification of brain states in rat neural recordings under different levels of anesthesia. To evaluate the robustness of our framework, we deliberately introduced noise artifacts into the neural recordings. We evaluated our hybrid Hopfield-CNN pipeline by benchmarking it against two comparative models: a standalone CNN handling the same noisy inputs, and another CNN trained and tested on artifact-free data. Performance across various levels of data compression and noise intensities showed that our framework can effectively mitigate artifacts, allowing the model to reach parity with the clean-data CNN at lower noise levels. Although this study mainly benefits small-scale experiments, the findings highlight the necessity for advanced deep learning and Hopfield Network models to improve scalability and robustness in diverse real-world settings.
\end{abstract}

\section{Introduction}

Brain states are specific configurations of neural activity characterized by distinct patterns of neuronal firing, spatiotemporal dynamics and behaviour \cite{harris2011}. These configurations arise in various behavioral and cognitive conditions and their understanding is fundamental to both neuroscience research and clinical practice \cite{gervasoni2004}, since they underpin the brain's ability to process information and interact with the external world. This knowledge holds significant implications for understanding both normal brain function and the treatment of neurological disorders \cite{rosanova2018}.

Natural brain states such as sleep versus wakefulness, or pharmacologically-induced brain states such as under anesthetics provide functional models for neuroscience research, as they enable the induction of specific spatiotemporal cortical activity patterns \cite{rosanova2018}. In this context, deep anesthesia is characterized by synchronous slow oscillations (SO) in the cortex \cite{sanchez-vives2017}. These SOs alternate between active 'Up states' and quiescent 'Down states,' offering simpler models for neurological investigation. Here we will concentrate in transitions involving SOs, wakefulness (AW), which shows desynchronized activity, and microarousals (MA) that are brief, transient shifts towards wakefulness that interrupt SO patterns. Understanding these global brain states—AW, SO, and MA - and the transitions between them is still a key challenge for neuroscience. A part of this challenge remains in the analysis required to identify these states efficiently, particularly due to the need for data compression strategies to handle large time-series datasets and effective artifact removal techniques to negate experimental noise \cite{li2015}.

In this study, we employ Convolutional Neural Networks (CNNs) and pre-processing via Hopfield Networks to improve the classification of  artifact-afflicted anesthesia brain states. Using low-dimensional binary images of high-Signal-to-Noise Ratio (SNR) neural channels, we explore trade-offs between accuracy, computational efficiency, and artifact distortion.

\section{Related Work}

\textbf{Computational Methods for Brain Data}.
Computational advances have revolutionized neural data analysis. While traditional statistical methods remain important for basic research \cite{tort-colet2021}, machine learning techniques like CNNs have gained prominence for tasks like anomaly detection and artifact-contaminated signal decoding \cite{li2015}. These methods also enable state classification through interpretable deep learning \cite{lee2023}. On the clinical side, breakthroughs have occurred in fields like robotics and brain-computers interfaces \cite{bonizzato2018}, as well as cutting-edge neuroprosthetics for speech decoding \cite{metzger2023}.

\textbf{Hopfield Networks and NeuroAI Applications}
Hopfield Networks, originally introduced by Hopfield in 1982 \cite{hopfield1982}, have gained renewed attention for their ability to act as associative memories in complex neural networks \cite{lucibello2023}. Modern variants of these networks, often trained via backpropagation, have shown superior memory storage properties and have been used as submodules in larger AI networks \cite{krotov2016}. Additionally, the interplay between associative memory models and neuroscience has been a burgeoning area of research, offering bi-directional insights that enrich both fields \cite{whittington2021hipo}, \cite{whittington2021}. Particularly relevant to this study are applications of Hopfield Networks for denoising and reconstructing neural data, showcasing the potential of these models to enhance our understanding of complex neural systems \cite{burns2023}\cite{bricken2021}.

\section{Methodology}

\begin{figure}[h]
    \centering
    \includegraphics[width=0.963\textwidth]{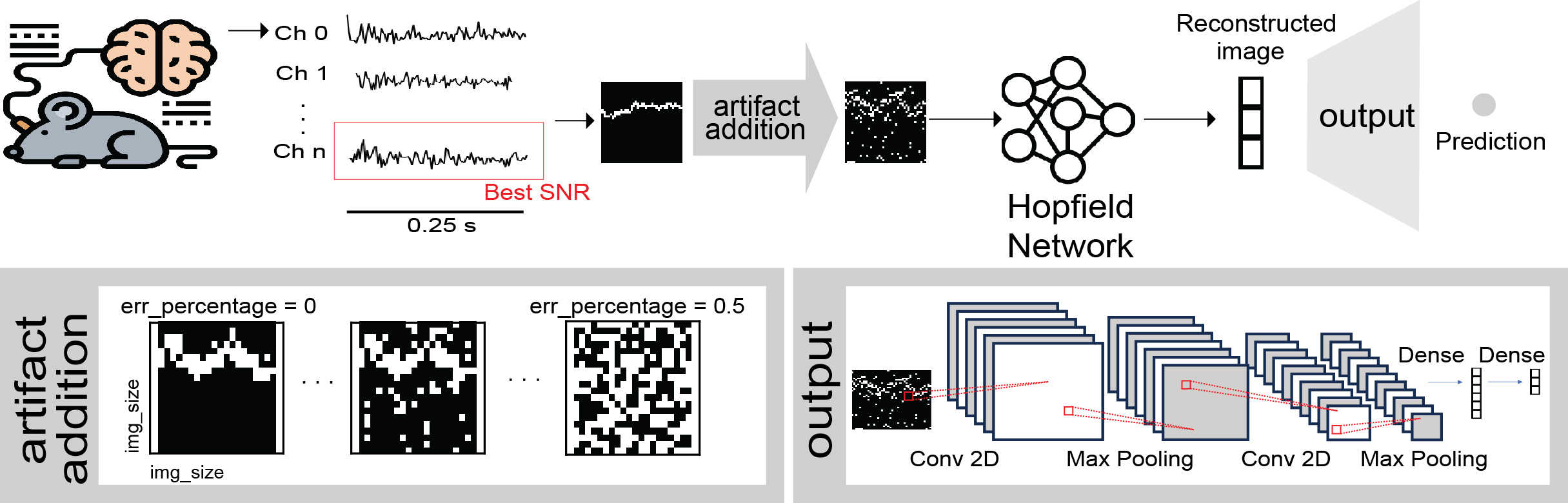}
    \caption{\textbf{Pipeline Overview}. Top row illustrates the general workflow of our methodology. Bottom row provides examples of various levels of artifact addition within a single input image, as well as the key layers involved in the model architecture for output generation.}
    \label{fig:method_overview}
\end{figure}

This study proposes a two-stage pipeline for classifying brain states under varying anesthesia levels in rats. We selected the neural channel with the highest SNR, transformed it into a lower-resolution binary image, and introduced artifacts for evaluation of classification robustness. Artifact correction is performed using a simple Hopfield Network, and final classification is done using a CNN. See Fig.\ref{fig:method_overview}.

\subsection{Data Reconstruction and Artifact Addition}

In our study, we analyze intracortical local field potentials (LFPs) from rats under different anesthesia levels, sampled at 10~kHz with manual labeling for each state. We segment the time-series data into 0.25-second intervals and transform them into binary images through binarization. We then adjust the image size by varying the compression ratio from \( \frac{1}{25} \) to \( \frac{4}{25} \). After binarization, we introduce artifacts by altering pixel values, guided by the \texttt{err\_percentage} parameter. For a more comprehensive explanation, see Appendix~A.

\subsection{Hopfield Network Model for Data Reconstruction}

To address the inherent capacity constraints of Hopfield Networks, we employ a multi-step approach. Firstly, a k-means clustering algorithm is used on the binary images for each of the states of anesthesia to select representative patterns. Each cluster's centroid serves as a representative pattern, thus allowing us to capture the underlying data structure while adhering to the \texttt{max\_patterns} limitation \cite{hopfield1982}. Next, individual Hopfield Networks are trained for each neural state using Hebbian learning. Specifically, the Hebbian rule is applied as \( w_{ij} = \sum_{p} x_i^{(p)} x_j^{(p)} \), where \( w_{ij} \) is the weight between neurons \( i \) and \( j \), and \( x^{(p)} \) is the \( p^{th} \) pattern. Note that for our implementation, self-connections \( w_{ii} \) are set to zero. These networks incorporate only artifact-added images, with the synaptic weights stored in \texttt{trained\_weights}. Finally, a template matching algorithm \cite{templatematching} is deployed for data reconstruction, leveraging Hamming distance as a similarity metric. The Hamming distance \( D_H \) between two images \( A \) and \( B \) is computed as \( D_H(A, B) = \sum_{i=1}^{n} |a_i - b_i| \), where \( n \) is the total number of pixels in the image, and \( a_i \) and \( b_i \) are individual pixels in \( A \) and \( B \) respectively. This enables the selection of the most similar clean image from the Hopfield Network for each artifact-corrupted image, thus completing the reconstruction process. 

\subsection{Convolutional Neural Network (CNN)}
The CNN takes artifact-corrected images as input, forming a supervised dataset that is labeled according to the originating neural states. The CNN, tailored for binary images, uses two convolutional and max pooling layers, followed by two dense layers with ReLU activations, ending in a softmax classification layer. Despite its smaller scale, the architecture balances resource efficiency with task-specific robustness. This design choice is corroborated by previous literature, for example as referenced in \cite{Tsinalis}. For more details on model training and optimization, please refer to Appendix A.

\subsection{Comparison and Validation of Results}

Assessing the efficacy of our hybrid model for brain state classification is complex due to an absence of standardized benchmarks. Traditional methods often rely on manual curation and visual assessment, lacking quantitative metrics for direct comparison. To tackle this, we compared the performance of our combined Hopfield Network and CNN approach against two alternative methodologies. To reduce variability, each model was run five times using different random seeds.

The first alternative employs an identical CNN model, but omits both the artifact introduction and the Hopfield Network reconstruction, feeding the CNN directly with the binary images. The second alternative includes artifact introduction but bypasses Hopfield Network-based reconstruction, providing the CNN with artifact-corrupted images for classification. These comparisons allow us to isolate the contributions of the artifact-handling elements in our hybrid model and gauge their impact on classification accuracy.

\section{Experimental Results}
\label{results}

\begin{figure}[ht]
\centering
\includegraphics[width=0.9\textwidth]{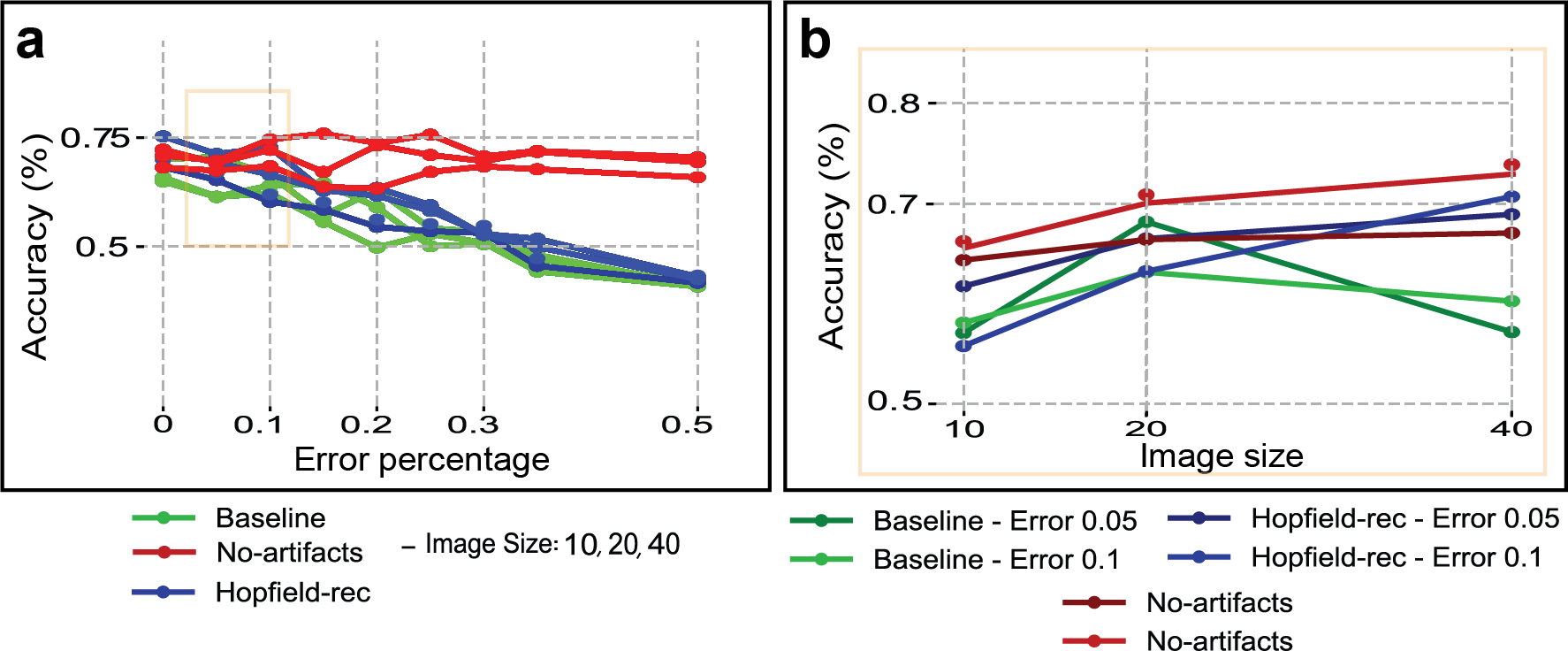}
\caption{\textbf{Experimental Results}: The figure assesses classification accuracy (\texttt{test\_acc}) using three strategies: CNN-only without artifacts (shown in \textcolor{Red}{red}), Hopfield Network + CNN (in \textcolor{Blue}{blue}), and a baseline incorporating artifacts (in \textcolor{Green}{green}). Subfigure \textbf{a)} explores how \texttt{test\_acc} varies with \texttt{err\_percentage}, aggregated over all image sizes. The accuracy of the CNN-only method without artifacts stays consistent across various error percentages, as it repeatedly processes identical inputs, akin to having different initialization seeds despite a constant zero error rate. Subfigure \textbf{b)} illustrates the relationship between \texttt{test\_acc} and image size, keeping \texttt{err\_percentage} range of 0.05 to 0.1 (highlighted in \textcolor{Apricot}{orange} squares).}

\label{fig:results}
\end{figure}

We employed Hopfield Networks for artifact removal in neural time series data, aimed at classifying transitions in brain states from deep anesthesia to wakefulness. This two-stage pipeline was benchmarked against two alternative methods: one that utilizes a neural model without artifact addition, using only clean signals; and a baseline strategy that retained artifacts but omitted Hopfield Network pre-processing. Our evaluation focuses on the efficacy of these various methods in brain state classification.

Figure \ref{fig:results}.A indicates that at low \texttt{err\_percentage} levels (0.05, 0.1), \texttt{test\_acc} is comparable between Hopfield Network reconstruction and CNN-only methods, especially at an image size of 40 where original frequencies are better preserved. At smaller image sizes (10, 20), less accurate frequency representation leads to similar \texttt{test\_acc} performance across all methods. For \texttt{err\_percentage} exceeding 0.2, the image distortions become too severe for effective reconstruction, aligning the \texttt{test\_acc} more closely with the baseline, as observed in Figure \ref{fig:results}.B.

\section{Discussion and Future Work}
\label{discussion}

We adopted a simple compression method for computational efficiency, at the expense of model accuracy. Future research should investigate advanced compression algorithms, such as autoencoders, to enhance state differentiation and frequency representation. Fourier analysis might be a promising avenue given its efficacy in capturing distinct frequencies characteristic of different anesthesia states like AW and SO \cite{manasanch}. Moreover, refining the model architecture could involve partitioning the MA states into asynchronous and synchronous components, as suggested by recent literature \cite{tort-colet2021}.

This study offers useful insights for small-scale projects but warrants further refinement for broader applications, particularly given challenges like inter-subject variability. Our results suggest the efficacy of Hopfield Networks for artifact reconstruction brain state classification under anesthesia, yet these findings are preliminary. Future work should scrutinize the robustness of this method across different distortion types and compression levels. Despite constraints on the number of patterns we can store, these limitations seem to have minimal impact on classification accuracy in small-scale experiments. We posit that the high similarity within short data segments of the same anesthesia state mitigates potential template-matching errors. However, comprehensive evaluation is needed to quantify these effects and uncover additional limitations and strengths.

\section{Availability}
All relevant code and demonstration dataset is publicly available at \url{https://github.com/arnaumarin/HDNN-ArtifactBrainState}.

\section{Acknowledgments and Disclosure of Funding}
This work was made possible by European Union’s Horizon 2020 Framework Programme for Research and Innovation under the Specific Grant Agreement No. 945539 (Human Brain Project SGA3) and by CORTICOMOD PID2020-112947RB-I00 funded by MCIN/ AEI /10.13039/501100011033.. We thank Leonardo Dalla Porta for his support, Melody Torao for obtaining originally the data, and we thank and acknowledge Cyprien Dcunha open source notebooks.

\section{Appendix A: Experimental Setup and Training Details}
\label{appendix A}

\textbf{Experimental Dataset:} Our demo-study focuses on approximately 5 minutes extracted from a 6-hour intracortical local field potential (LFP) recording from a rat under ketamine-medetomidine anesthesia. Each anesthesia state is adequately represented. The dataset is sampled at a frequency of 10 kHz and manually labeled by experts for ground truth annotations. The intracortical LFP time-series is divided into 0.25-second segments, chosen to capture stationary features across different anesthesia states. These segments are transformed into binary images using a custom binarization algorithm pipeline. 

In this pipeline, each 0.25-second segment is first normalized based on a globally computed minimum and maximum value, ensuring consistent intensity scaling across different states. Each normalized time-value is then mapped to a pixel coordinate in the binary image; the x-coordinate corresponds to the time index, and the y-coordinate to the normalized value. This transformation effectively spatializes the time-series data into a 2D binary image, capturing the waveform shapes inherent to the original series. To explore the impact of compression, we manipulate the \texttt{image\_size} between 10 and 40, corresponding to compression ratios of 1/25 and 4/25, respectively. This image-based representation serves as a practical and simple medium for neural data analysis, allowing for varying degrees of compression.

After the binarization step, we introduce artificial artifacts by algorithmically altering a predetermined subset of pixels. This subset is determined by the \texttt{err\_percentage} parameter, which serves to emulate common neural recording noise and artifacts \cite{artifacts_types}. The \texttt{err\_percentage} values can range between 0 and 0.5, signifying that up to 50\% of the total pixels may be corrupted.

\begin{figure}[h]
    \centering
    \includegraphics[width=1\textwidth]{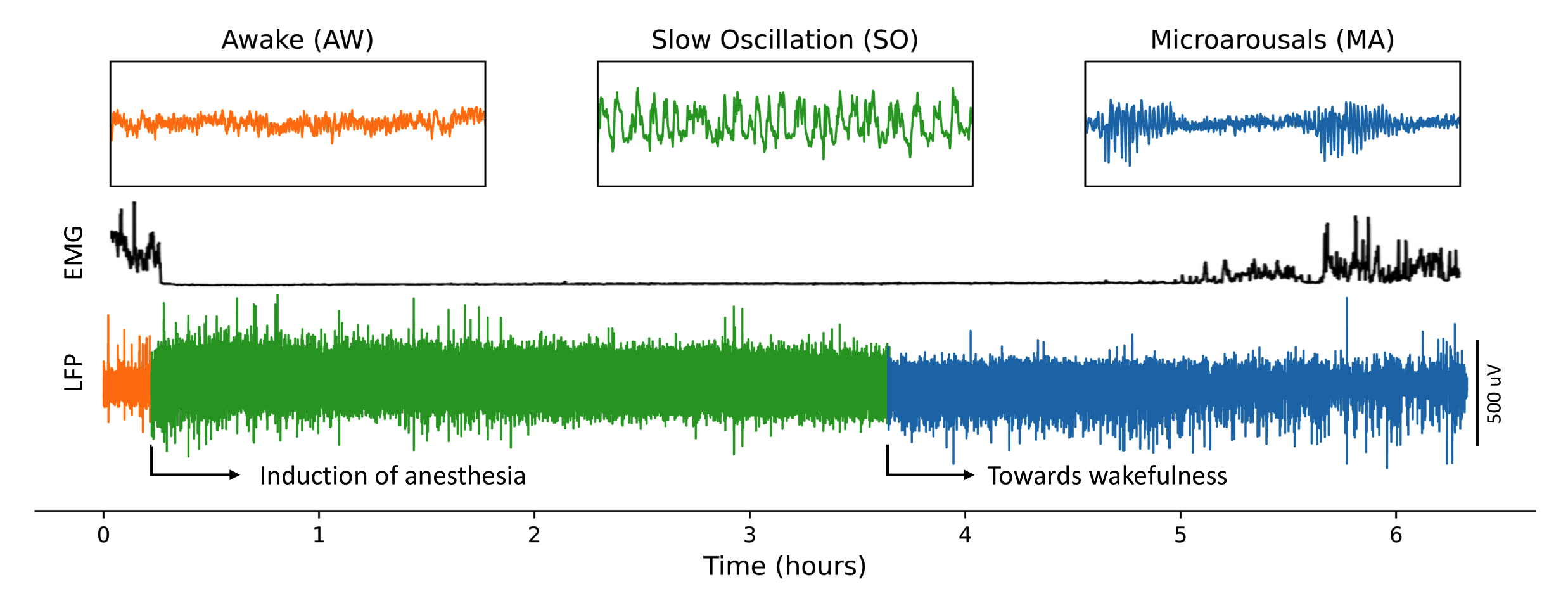}
    \caption{\textbf{Schematic of the different Ground Truth brain states}. LFP of a full transition is shown. It starts at Awake (\textcolor{Orange}{orange}) and after the anesthesia induction it rapidly falls into the SO (\textcolor{Green}{green}) regime. After a few hours it goes to the MA state (\textcolor{Blue}{blue}), approaching wakefulness, as can be seen by the EMG amplitude.}
    \label{fig3 GT}
\end{figure}

\textbf{Ground truth annotations:}
Manual labeling for the establishment of ground truth annotations was carried out by an expert who was instrumental in data collection. This process involved defining three significant global states stated before: AW, SO, and MA. Ground truth annotations were applied to the LFP channel exhibiting the highest SNR. Additionally, an electromyography (EMG) channel, positioned around the animal's neck, was utilized to monitor animal movements and served as a supplementary source of information for this labeling process. This dataset originates from a series of experiments designed to assess various brain states occurring within the rat cortical brain under the influence of ketamine and medetomidine anesthesia. Each experiment spanned approximately 5-6 hours, as shown in Figure \ref{fig3 GT}. The entire recording was categorized as follows.

The AW state pertained to time periods showing asynchronous cortical activity recorded before anesthesia administration. During this phase, the animal exhibited clear signs of wakefulness, such as responsiveness and movement, accompanied by a high EMG amplitude. SO labels were assigned to time segments characterized by a distinct Up/Down pattern with a frequency below 2.5 Hz. These segments immediately followed anesthesia induction and typically persisted for 2-3 hours. The animal displayed no signs of wakefulness, and the EMG exhibited a flat profile. MA labels were designated for time samples that occurred once the stable SO state had concluded. These samples were expected to exhibit a shift toward wakefulness, as detailed in \cite{tort-colet2021}. They consisted of brief periods of asynchronous activity interspersed with intervals of slow oscillation characterized by a notably higher frequency (>2.5 Hz) compared to the typical SO observed after anesthesia induction. During the MA state, the animal remained unresponsive but began to display drowsiness-related symptoms, which were validated through EMG and video recordings.

Any time samples showing noise or the presence of electrical or movement artifacts were categorized as "artifact". Samples where definitive decisions could not be reached were classified as "unknown".

\textbf{CNN Model Training and Optimization:} 

The CNN takes artifact-corrected images as input, forming a supervised dataset that is labeled according to the originating neural states. Engineered to operate on binary images, the CNN architecture incorporates two 2D convolutional layers with a kernel size of 3, each succeeded by max pooling layers. The architecture concludes with two fully connected dense layers that use ReLU activations, except for the softmax-activated final layer dedicated to classification. It should be noted that although our architecture is smaller compared to typical benchmarks in the field, it was designed to be resource-efficient while still being sufficiently robust for the tasks at hand.

The Convolutional Neural Network (CNN) model is subjected to a grid search to optimize the number of epochs and the learning rate; the options considered are [5, 10, 15], and [0.1, 0.001, 0.0001] respectfully. It is important to note that we did not optimize the other hyperparameters because the primary goal of this study is not to maximize and find the optimal classification performance. Instead, we focus on assessing the Hopfield network's ability to reconstruct signals affected by artifacts.

Regarding dataset partitioning, 10\% of the data is allocated for training, and another 10\% is set aside as a validation set to monitor the model and prevent overfitting. The remaining 80\% of the data serves as the test set. To evaluate the effectiveness of the model, we use test accuracy, denoted as \texttt{test\_acc}, and a normalized confusion matrix, referred to as \texttt{cm\_normalized}, as our key performance indicators.

\end{document}